# Ordered (3×4) High Density Phase of Methylthiolate on Au(111)

*Valentina De Renzi,\*† Diego Marchetto,† Roberto Biagi,† Umberto del Pennino,† Rosa Di Felice,†*

*Annabella Selloni‡*

†INFM Center for nanoStructures and bioSystems at Surfaces ($S^3$) and Dipartimento di Fisica,
Università di Modena e Reggio Emilia, Modena, Italy;

‡Department of Chemistry, Princeton University, Princeton, New Jersey, NJ 08544

## ABSTRACT

The formation of ordered phases of dimethyl-disulfide on the Au(111) surface has been investigated by means of Low-Energy Electron Diffraction (LEED), X-ray Photoemission Spectroscopy (XPS), and state-of-the-art Density Functional Theory (DFT) periodic supercell calculations. The LEED diffraction pattern, obtained after a production method that includes a two-step dosing and prolonged post-deposition annealing, unambiguously corresponds to a novel phase that consists of (3×4) domains coexisting with the as-deposited ($\sqrt{3}\times\sqrt{3}$)R30° structure. XPS measurements indicate that the coverage of the new (3×4) superstructure is the same as that of the ($\sqrt{3}\times\sqrt{3}$)R30° phase. In both phases, the binding energy of the S $2p_{3/2}$ core-level peak is found to be 162.2 eV, corresponding to the formation of a thiolate layer. The DFT calculations allow us to identify a viable metastable (3×4) structure where the S headgroups of the CH$_3$S radicals select distinct adsorption sites: three quarters of them adsorb at bridge sites and one quarter at top sites. The relative energetics of the (3×4) and ($\sqrt{3}\times\sqrt{3}$)R30° configurations suggest that the two structures may coexist on the surface, in agreement with experimental data.

---

\* corresponding author: e-mail vderenzi@unimore.it  fax +39-059-2055235

**Introduction**

Interfaces between organic molecules with sulfur-containing headgroups and noble metal surfaces represent a fruitful class of systems with potential relevance in new technological applications, such as supramolecular assembly, wetting and corrosion inhibition[1]. In particular, the capabilities of S headgroups to strongly bind to different metals is exploited in state-of-the-art processing to obtain functionalized surfaces[2] as well as templates for further organic-on-organic growth[3,4]. Despite alkanethiols CH$_3$(CH$_2$)$_{n-1}$SH (Cn) on Au(111) are the best studied systems among this class of self-assembled monolayers (SAMs), several questions are still open, concerning their adsorption mechanism and the relevance of inter-molecular interactions in driving their two-dimensional order. These are fundamental issues to be addressed for the achievement of a full control on the structural and electronic properties of functionalized surfaces[5].

One relevant point still to be elucidated is the role of the thiol chain length in determining the molecular self-assembly at full coverage. Long-chain alkanethiols have been shown to form a ($3\times 2\sqrt{3}$) rectangular superlattice at full coverage[6,7,8], in a wide range of preparation conditions including deposition from solution and from vapor, with and without post-deposition annealing Whereas a sulfur-pairing model implying multiple S-Au interactions was initially proposed for the ($3\times 2\sqrt{3}$) unit cell in the case of C10 and C8 on Au(111)[8,9,10], recent high resolution X-ray spectroscopy (XPS) studies of similar ordered decanethiol monolayers[11] reported a single S 2p$_{3/2}$ core level peak, suggesting a unique thiolate-gold interaction mechanism for all the thiols in the unit cell. Therefore, it is still a matter of debate whether the $(3\times 2\sqrt{3})$ phase is mainly due to layer-substrate coupling, or should rather be ascribed to lateral chain-chain interactions possibly dictated by the existence of different S binding sites but with the same S-Au coupling.

Studies of shorter thiols where the chain-chain interactions are minimized were thus conceived to elucidate the above issue, and several ordered phases were reported. Focusing our attention on the shortest possible alkanethiol chains with n=1, we note in particular that Dishner et al.[12] observed by Scanning Tunneling Microscopy (STM) the formation of both a $(3\times 2\sqrt{3})$ structure and a $(2\sqrt{3}\times\sqrt{3})R30^o$ striped phase for methanethiol (C1, *e.g.* $CH_3SH$) SAMs on Au(111), at full and partial monolayer coverage, respectively. Recently, a He-Atom Scattering (HAS) experiment found a new ordered phase obtained from dissociative adsorption of dimethyl-disulfide $(CH_3S)_2$ on Au(111) after a proper post-deposition annealing procedure, which has been interpreted as a $(3\times 2\sqrt{3})$ rectangular reconstruction[13] with the same periodicity as that found for the longer chains. On the basis of this result, the authors suggested that the $(3\times 2\sqrt{3})$ phase is energetically favored for all alkanethiols, independently of the chain length, and that its attainment is simply determined by a proper sample preparation. The existence of a complex phase diagram also in the cases where the alkylic chains are short and lateral coupling is expected to play a minor role in the monolayer aggregation, may be interpreted as a signature of the fact that the complex reconstructions are prominently induced by S-Au interactions, which involve geometrically inequivalent but electronically similar binding sites[11], characterized by similar adsorption energies and thiolate coupling strengths.

From the theoretical point of view, DFT-based quantum mechanical computational techniques are suitable to characterize the surface superstructure induced by S-metal coupling but lack an accurate description of the inter-chain coupling: thus, they were mainly employed to describe short-chain thiols (in particular $CH_3S$, where the tailgroups are expected to play a minor role) on different surfaces[14,15,16]. On the contrary, classical molecular dynamics and molecular mechanics have been the methods of choice to take into account the weak dispersion-like interactions between the molecular tailgroups of long-chain alkanethiols[17,18,19], but give a poor description of the S-metal interface bond. As a consequence, a direct energetical comparison of the two sets of alkanethiols (short-chain *versus* long-chain) is impossible on the basis of the available results and with the state-of-the-art computational settings. A combination of the quantum and the classical approaches may be the key to solve this long-standing controversy: An application of such a coupled quantum-mechanics/molecular-mechanics (QM/MM) scheme to the study of C10/Au(111) monolayers has only quite recently appeared[20], shedding light onto several issues about the structure and the energetics of long-chain alkanethiol monolayers on metal surfaces. Whereas QM/MM computations are promising candidates for a solution of the long-standing controversies about the self-assembly of alkanethiol monolayers on metal surfaces, DFT calculations based on purely QM schemes currently remain a reliable and more affordable tool to investigate the interface features of short-chain thiolate layers.

Recent DFT calculations on the $CH_3S/Au(111)$ system found the $(3 \times 2\sqrt{3})$ phase energetically indistinguishable from the $(\sqrt{3} \times \sqrt{3})R30^o$ hexagonal lattice[21]. Despite this result is not able to explain the observation of the $(3 \times 2\sqrt{3})$ superstructure on the basis of a purely energetical point of view, it supports the existence of this phase as an equilibrium structure, and confirms that its occurrence in the reality may be stimulated by the proper deposition technique. In a similar way, DFT calculations may be employed to analyze the geometry and the energetics of other surface reconstructions.

In this letter, we report the results of further experimental and theoretical investigations on the formation of ordered thiolate phases obtained by deposition of dimethyl-disulfide (DMDS) on the Au(111) surface. Following the recently proposed preparation procedure[13], we observe by Low-Energy Electron Diffraction (LEED) the formation of a highly ordered (3×4) phase. In fact, the observed LEED pattern, which turns out to be fully compatible with the observations by Danisman et al.[13], corresponds to the coexistence of both $(\sqrt{3} \times \sqrt{3})R30^o$ and (3×4) phases rather than the proposed $(3 \times 2\sqrt{3})$ phase. On the basis of DFT calculations, we propose a viable model for the observed (3×4) reconstruction, whose molecular fractional coverage is the same as that of the $(\sqrt{3} \times \sqrt{3})R30^o$ phase, in agreement with the indications of our XPS results. The calculated structure corresponds to a metastable phase, and the energetic relationship indicates that it is likely to coexist with the ($\sqrt{3}\times\sqrt{3}$)R30° order[9,21].

**Results and Discussion**

The experiment is performed in a UHV chamber (base pressure $1\times10^{-10}$ mbar), where sputtering facilities, X-ray photoemission and LEED equipments are available. The X-ray source (Mg $k_\alpha$, photon energy hν=1253.6 eV) is non-monochromatized and the overall energy resolution is 2 eV. The sample is mounted on a cryogenic manipulator and all measurements are performed at 100 K, if not otherwise stated. After cleaning by sputtering-annealing cycles, the sample is exposed to DMDS at 200 K. The DMDS exposure is performed by back-filling the chamber through a leak-valve (typical dose pressure $4\times10^{-7}$ mbar). Previous to every exposure DMDS is purified by several freeze-pump-thaw cycles. Organic molecules are known to be significantly damaged by electron beam irradiation. For this reason, we minimize the surface exposure to the LEED electron beam, maintaining the exposure time shorter than 1 minute, with a beam current density estimated about 50 nA/mm².

DMDS is expected to dissociate at room-temperature (RT) on the Au(111) surface, forming a thiolate layer[14,22,23]. In our experiment, DMDS is dosed on the clean Au(111) surface at 200 K and subsequently annealed to 320 K[13]. Exposure to 33 L DMDS results in the formation of the well known ($\sqrt{3}\times\sqrt{3}$)R30° phase, as shown in the LEED pattern reported in Fig. 1a. Interestingly, at lower coverage the LEED pattern reveals, in addition to the spots of the ($\sqrt{3}\times\sqrt{3}$)R30° superstructure, the presence of weak satellites of the integer-order peaks and some streaky features connecting both integer- and fractional-order spots (see Fig. 1b). This pattern can be tentatively assigned to the formation of small islands with different (n$\sqrt{3}\times\sqrt{3}$)R30° striped phases, the position of the satellite spots fitting in particular with n=4. In Fig. 1c, the calculated LEED pattern for a (4$\sqrt{3}\times\sqrt{3}$)R30° superstructure is shown for comparison. This finding is in substantial agreement with the STM observation of a striped phase at low coverage[12], which is determined by missing rows of molecules in the $(\sqrt{3}\times\sqrt{3})R30°$ lattice.

Following Ref.[13], the well-ordered $(\sqrt{3}\times\sqrt{3})R30°$ phase – obtained by exposing a freshly prepared Au(111) surface to 33L DMDS and annealing to 320K – is then further exposed to 300 L DMDS, while ramping the temperature between 200 and 273 K. The sample is subsequently annealed at RT overnight, eventually obtaining the LEED pattern shown in Fig. 2a. This pattern is due to the coexistence of the $(\sqrt{3}\times\sqrt{3})R30°$ phase with a new (3×4) phase, as shown schematically in Fig. 2b[24]. For the sake of clarity, we refer in the following to this complex structure as *coexisting-(3´4)* phase. It is important here to notice that the observed LEED pattern does not fit with that of a $(3\times 2\sqrt{3})$ superstructure. On the other hand, it compares quite well with the HAS contour plot by Danisman and coworkers[13]. Due to the relatively low resolution of the HAS apparatus used in ref.[13], their data cannot discriminate between the two structures. The average fractional coverage of the *coexisting-(3´4)* phase, as determined by XPS measurements, is the same as that of the $(\sqrt{3}\times\sqrt{3})R30°$ structure. Moreover, the shape and energy position of the S 2p core-level peak (not shown) is the same for both phases and corresponds to a thiolate layer (S $2p_{3/2}$ at 162.2 eV binding energy). A small component at 164.1 eV is observed in both spectra, which may be attributed to a small fraction of molecules not directly bound to the metal surface[25]. No atomic sulfur contribution is detected at the lower binding energy side of the S 2p spectra.

The presence of local (3×4) ordered structures on the $CH_3S$/Au(111) system was already observed by STM at low temperature by Kondoh and Nozoye[26], who suggested that a phase transition from the (3×4) to the $(\sqrt{3}\times\sqrt{3})R30°$ phase would occur upon passing from low temperature (110 K) to room temperature (RT). In order to check this indication, we investigate the temperature dependence of the LEED pattern of both the $(\sqrt{3}\times\sqrt{3})R30°$ and the *coexisting-(3´4)* phases between 100 K and 330K. In the case of the $(\sqrt{3}\times\sqrt{3})R30°$ phase, no structural change can be observed by LEED in this temperature range. On the other hand, the *coexisting-(3´4)* phase displays a strong temperature dependence: (i) A reversible disappearance of the (3×4) spots around 310 K may tentatively be interpreted as the fingerprint of a first-order (3×4)-to-($\sqrt{3}\times\sqrt{3}$)R30° phase transition, in partial agreement with the STM findings; (ii) Upon annealing to 330 K the *coexisting-(3´4)* phase irreversibly changes into the ($\sqrt{3}\times\sqrt{3}$)R30° phase, in agreement with previous experimental reports[13] and with theoretical indications of a superior stability of the latter long-range order[21]. A thorough temperature-dependent characterization of the *coexisting-(3´4)* structure is currently the object of further investigations.

To complement the experimental data and support the existence of a (3×4) reconstruction for the thiolated Au(111) surface, we perform ab-initio plane-wave pseudopotential DFT-PW91 periodic slab calculations[27]. The positions of all the atoms in the supercell are relaxed in the potential energy determined by the full quantum mechanical electronic structure, until the forces vanish within a precision of 0.03 eV/Å. The electron-ion interaction in the DFT total energy functional is described by non-norm-conserving pseudopotentials[28] for the species C, H, Au, whereas S was represented by a norm-conserving pseudopotential[29]. The electron wavefunctions are expanded in a plane-wave basis-set up to a kinetic energy cutoff of 20 Ryd. The surface is simulated with a supercell having a 3×4 periodicity, containing 4 Au layers with 12 atoms each, 4 $CH_3S$ molecules adsorbed at one surface in the slab, and a vacuum thickness of 12 Å to avoid spurious interactions between neighboring replicas (see Fig. 3). Four special k points are included in Brillouin Zone (BZ) sums. The computational details were previously validated by tests on similar systems[21,23].

Our calculations allow to identify a metastable (3×4) structure, shown in Fig. 3, where the S headgroups of the $CH_3S$ radicals select distinct adsorption sites: three quarters of them adsorb at bridge sites B1 and B2 (including bridging directions along both the independent basis vectors of the hexagonal lattice), and one quarter at top sites T of the (111) *fcc* lattice. The thiols are distributed in stripes which are approximately parallel to the major diagonal. Along such stripes, there are two inequivalent S-S distances equal to 4.25 Å and 4.83 Å, for the bridge-top and the bridge-bridge pairs, respectively. The C-S bond distance is 1.87 Å for all the thiols in the unit cell, whereas the inclination of such bonds with respect to the surface normal varies between 42° and 58° for the molecules with the S headgroups anchored at B sites, and is 64° for those at T sites. The average S-Au distance is 2.5 Å (2.4 Å at the T site), typical of thiolate-gold interfaces independently of the detailed reconstruction and even of the molecular adsorbate: for instance, the S-Au distance is the same when cysteine is the adsorbed molecule[23]. The topmost Au layer is strongly corrugated, by about 0.3 Å. To support the viability of the optimized (3×4) geometry, we compare its total energy with that of the ($\sqrt{3} \times \sqrt{3}$)R30° structure: for the latter, we choose the lowest-energy configuration[21] with all the S headgroups adsorbed at bridge sites and thus equally spaced by 5.1 Å, a S-Au distance of 2.5 Å, the S-C bonds forming an angle of 56° with respect to the surface normal, and a topmost Au corrugation of 0.3 Å. We find that the computed (3×4) geometry is slightly energetically unfavored relative to the ($\sqrt{3} \times \sqrt{3}$)R30° phase by about 1 kcal/mol[30], in agreement with the experimental observation of the *coexisting-(3´4)* phase. We note that both the occurrence of two different bridge sites (see B1 and B2 in Fig. 3) and the energetic balance between the (3×4) extended-cell superstructure and the basic hexagonal ($\sqrt{3} \times \sqrt{3}$)R30° geometry, are in line with similar results of the recent QM/MM computations for the ($3 \times 2\sqrt{3}$) monolayers of decanethiols on Au(111)[20]. Although the structures and the molecules are different in the two cases, the similarities suggest common effects due to the S-Au interactions. Whereas it is not possible at the adopted level of theory to discriminate between the relative importance of molecule-substrate and molecule-molecule coupling in the determination of complex reconstructions for C1/Au(111), the present calculations support the suggestion that a complex energy landscape and phase diagram of thiolated surfaces is characteristic also of short-chain alkanethiols.

**Conclusions**

In conclusion, we unambiguously determine a novel *coexisting-(3´4)* long-range ordered phase of the $CH_3S$/Au(111) system, obtained after a recently proposed dosing-annealing procedure[13] suitable to produce ordered thiolate monolayers on Au(111) from vapor-phase DMDS. This phase is characterized by the coexistence of the well-known ($\sqrt{3}\times\sqrt{3}$)R30° with a new (3×4) superstructure. Preliminary indications of a structural phase transition around 310 K are inferred by the analysis of the LEED temperature dependence. LEED measurements also indicate that at low coverage a striped phase is formed, characterized by small islands of (n$\sqrt{3}\times\sqrt{3}$)R30° order, in agreement with STM observations[12]. A viable theoretical model for the (3×4) superstructure is proposed, which is found to be metastable and with a formation energy higher by only 1 kcal/mol with respect to the ($\sqrt{3}\times\sqrt{3}$) hexagonal phase, thus justifying their coexistence. Our results therefore confirm, in agreement with HAS measurements[13], that different ordered phases are likely to occur at full coverage also for the smallest thiolate precursor. In the case of the $CH_3S$/Au(111), the observation of a long-range ordered (3×4) phase different from the (3×2$\sqrt{3}$) superstructure suggests that the molecular chain length plays a significant role in the determination of the thiolate geometrical details. Interestingly, a local (3×4) phase has been also observed with STM for ethanethiol layers on Au(111)[31], thus suggesting that this superstructure could be a common feature of short-chain alkanethiolate overlayers on the Au(111) surface. Finally, we would like to remark that the availability of different ordered superstructures of thiolate interfaces could be possibly exploited in soft-on-soft epitaxy to tailor the template geometrical properties.

**Acknowledgement**

M. Borsari and co-workers are acknowledged for experimental support. The authors are grateful to L. Casalis and G. Scoles for useful discussions and for sharing unpublished details of their data. Computing time at CINECA is provided by INFM through the Parallel Computing Committee. Funding is provided by INFM through PRA SINPROT, by MIUR (Italy) through FIRB NOMADE, and by the EC through contract IST-2000-28024 "SAMBA".

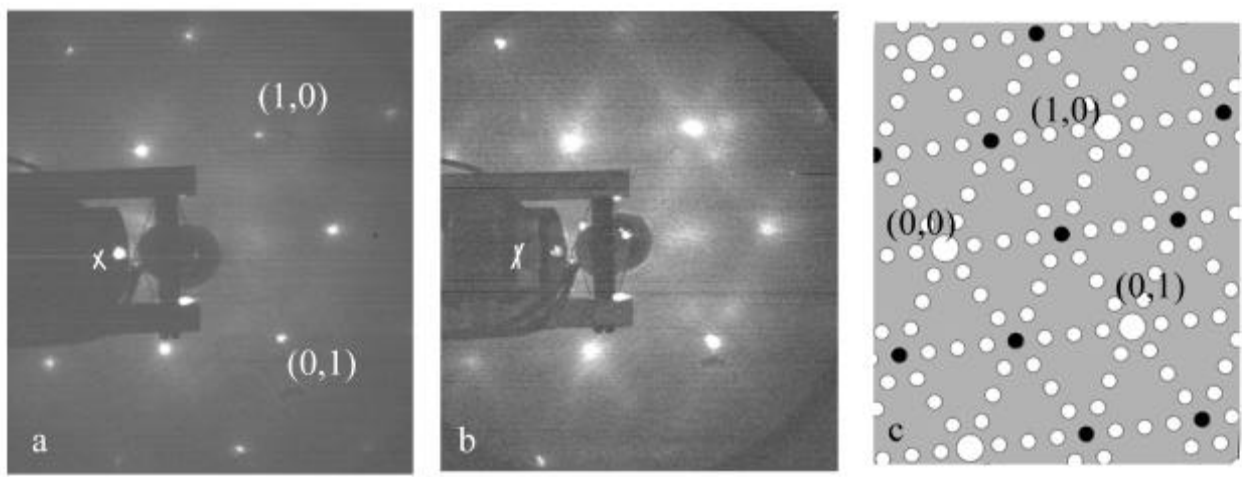

**Figure 1.** (a) LEED pattern of the ($\sqrt{3}\times\sqrt{3}$)R30° phase, obtained by dosing 33L DMDS at 200K and subsequently annealing to 320K for 2 minutes, taken at 100K, with electron beam energy $E_p$=60 eV. The white cross indicates the specular peak position. (b) LEED pattern of the low-coverage phase, obtained after dosing 3L DMDS at 100K and subsequent annealing to 320K for 2 minutes. XPS measurements on this surface indicated a fractional coverage of 0.66 relative to the ($\sqrt{3}\times\sqrt{3}$)R30° phase. (c) Scheme of the diffraction pattern of the ($4\sqrt{3}\times\sqrt{3}$)R30° superstructure: Large white circles represent integer-order spots, small black and white circles correspond to the ($\sqrt{3}\times\sqrt{3}$)R30° and ($4\sqrt{3}\times\sqrt{3}$)R30° spots, respectively.

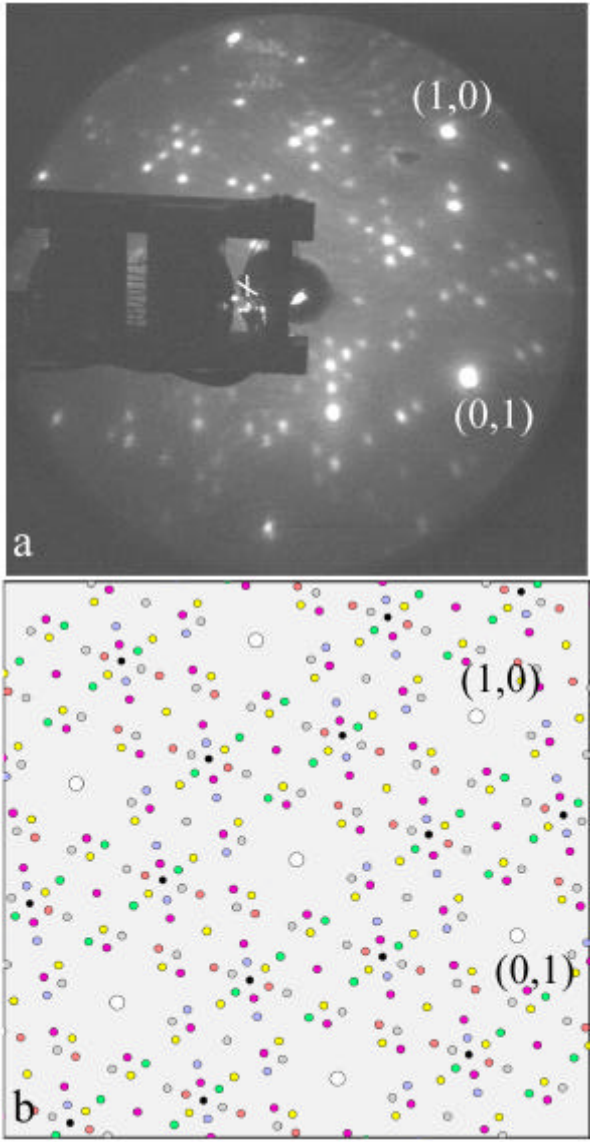

**Figure 2.** (a) LEED pattern of the *coexisting-(3´4)* phase, taken at 100K with $E_p$=60 eV. The specular beam position (white cross) and two integer order peaks are indicated. (b) Scheme of the diffraction patterns of the six equivalent domains of the (3×4) superstructure (small colored circles) and of the ($\sqrt{3}\times\sqrt{3}$)R30° superstructure (small black dots). White large circles corresponds to the integer order peaks.

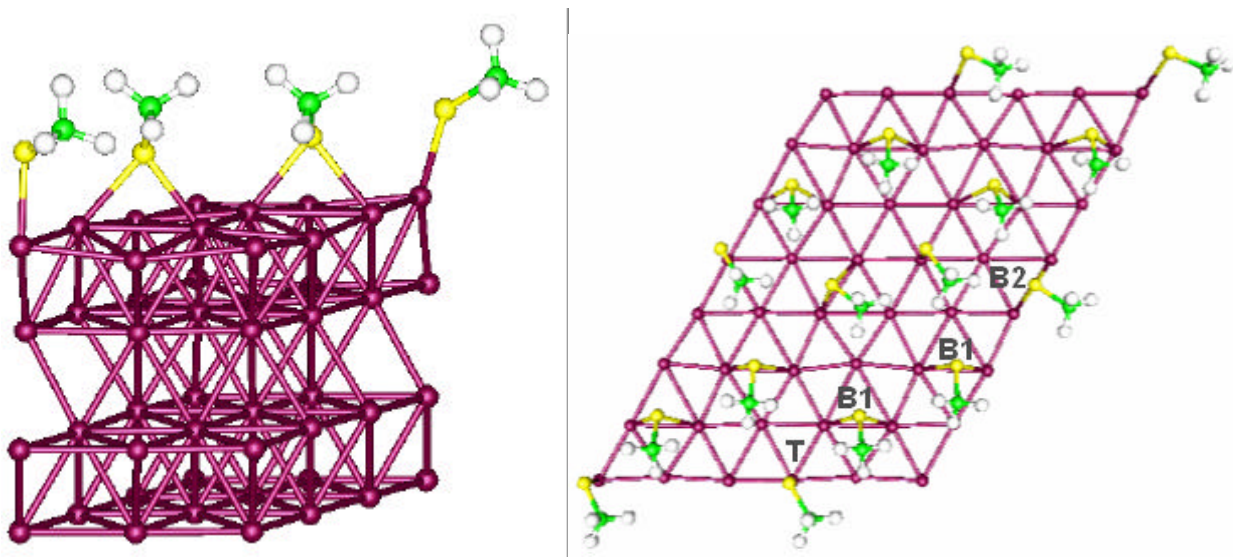

**Figure 3.** Left: three-dimensional view of the equilibrium structure. Right: top-view of the equilibrium structure where it is clear that ¾ thiols adsorb at bridge sites (½ B1 along lattice vector $a_1$, ¼ B2 along lattice vector $a_2$) ad ¼ thiols adsorb at top sites (T), forming stripes along the major diagonal of the unit cell. Purple, yellow, green, and white spheres represent Au, S, C, and H atoms.